\documentclass[12pt,a4paper]{article}

\usepackage{geometry}
\usepackage{setspace}
\usepackage{hyperref}
\usepackage{amsmath}
\usepackage{mathtools}
\usepackage{amssymb}
\usepackage{cite}

\usepackage{microtype}     
\usepackage{amssymb} 

\begin{document}

\begin{center}
{\Large \textbf{Contractions of the relativistic quantum LCT group and the emergence of spacetime symmetries}}\\[1.5cm]
\end{center}

\begin{center}
    \noindent
    \textbf{Anjary Feno Hasina Rasamimanana}$^{1}$, 
    \textbf{Ravo Tokiniaina Ranaivoson}$^{2}$, 
    \textbf{Roland Raboanary}$^{3}$, 
    \textbf{Raoelina Andriambololona}$^{4}$,
    \textbf{Wilfrid Chrysante Solofoarisina}$^{5}$,
    \textbf{Philippe Manjakasoa Randriantsoa}$^{6}$\\[1cm]
\end{center}

\begin{center}
    \noindent
    \texttt{anjaryhasinaetoile@gmail.com}$^{1}$,
    \texttt{tokiniainaravor13@gmail.com}$^{2}$,
    \texttt{r\_raboanary@yahoo.fr}$^{3}$,
    \texttt{raoelina.andriambololona@gmail.com}$^{4}$,
    \texttt{wilfridc\_solofoarisina@yahoo.fr}$^{5}$,
    \texttt{njakarandriantsoa@gmail.com}$^{6}$\\[1cm]
\end{center}


\begin{center}
   \noindent
    $^{1,2,4,5,6}$\textit{Theoretical Physics Department},
    \textit{Institut National des Sciences et Techniques Nucléaires (INSTN-Madagascar)},
    BP 3907 Antananarivo 101, Madagascar, \texttt{instn@moov.mg} \\[1cm]
\end{center}

\begin{center}
   $^{2,4}$\textit{TWAS Madagascar Chapter, Malagasy Academy},
    BP 4279 Antananarivo 101, Madagascar \\[1cm]
\end{center}

\begin{center}
   $^{1,3,6}$\textit{Faculty of Sciences, iHEPMAD-University of Antananarivo},
    BP 566 Antananarivo 101, Madagascar \\[1cm]
\end{center}

\begin{abstract}
Advances in the study of relativistic quantum phase space have established the set of Linear Canonical Transformations (LCTs) as a candidate for the fundamental symmetry group associated with relativistic quantum physics. In this framework, for a spacetime of signature $(N_+,N_-)$, the symmetry of the relativistic quantum phase space is described by the LCT group, isomorphic to the symplectic Lie group $Sp(2N_+,2N_-)$, which preserves the canonical commutation relations (CCRs) and treats spacetime coordinates and momenta operators on an equal footing. In this work, we investigate the contraction structure of the Lie algebra associated with the LCT group for signature $(1,4)$, clarifying how familiar spacetime symmetry groups emerge from this more fundamental quantum phase space symmetry. Using the Inönü-Wigner group contraction formalism, we examine each limit case corresponding to the possible combinations of asymptotic values of two fundamental length scale parameters associated with the theory, namely a minimum length $\ell$ and a maximum length $L$, which may be identified respectively with the Planck length and the de Sitter radius. We explicitly analyze how contractions of the LCT Lie algebra lead to the physically relevant de Sitter algebra $\mathfrak{so}(1,4)$ and, in the flat-curvature limit, to the Poincaré algebra $\mathfrak{iso}(1,3)$ of four-dimensional spacetime. This provides an explicit mechanism through which relativistic spacetime symmetry can emerge from a deeper symplectic structure of quantum phase space.
\end{abstract}


\section{Introduction}

LCTs are fundamental mathematical tools that appear in signal processing, optics, and quantum physics, providing a generalization of integral transforms such as the Fourier transform and its fractional counterpart\cite{Ranaivoson2021,Wolf1979,Xu2013,Healy2016}. 
In relativistic quantum physics, they act linearly on coordinates and momenta operators while preserving the CCRs. Since these commutation relations encode the fundamental structure of relativistic quantum theory, LCTs may naturally be interpreted as symmetry transformations of the relativistic quantum phase space \cite{Ranaivoson2021,Ranaivoson2022,Ranaivoson2025,Randriantsoa2025}.

In a recent study, it was shown that for a spacetime of signature $(N_{+},N_{-})$, the group of relativistic LCTs is isomorphic to the symplectic group $Sp(2N_{+},2N_{-})$  \cite{Ranaivoson2021,Ranaivoson2022,Ranaivoson2025,Randriantsoa2025}.
Although the symplectic structure of relativistic quantum phase space has been previously established in \cite{Ranaivoson2021}, the contraction structure of the corresponding Lie algebra and the emergence of spacetime symmetries from this framework have not been systematically analyzed. The purpose of the present work is to fill this gap.

An important conceptual question then arises: how are the familiar symmetry groups of spacetime related to this more general phase-space symmetry? 
Classical non-relativistic mechanics is covariant under the Galilei group \cite{Leblond1971}, special relativity under the Poincar\'e group, and spacetime with positive constant curvature under the de Sitter group \cite{Aldrovandi2007}. 
These groups describe different physical regimes of spacetime geometry. 
The LCT symmetry suggests that they may not be understood as independent starting points, but as effective limits of a deeper quantum symplectic structure \cite{Ranaivoson2021}. In particular, the de Sitter group may be viewed as corresponding to a regime in which some quantum and gravitational effects are neglected but the spacetime curvature is finite, while the Poincaré group arises in the flat-curvature limit. The Galilei group can  be recovered as the non-relativistic limit of Poincaré symmetry \cite{Wigner1953}. From this perspective, the familiar spacetime symmetry groups appear as successive physical limits of a more fundamental symmetry acting in quantum phase space.

In this paper, we make this emergence mechanism explicit by performing a systematic analysis of the contraction structure of the relativistic LCT algebra for signature $(1,4)$. Introducing two fundamental length scales, a minimum length $\ell$ and a maximum length $L$, we show how the de Sitter algebra $\mathfrak{so}(1,4)$ and the Poincaré algebra $\mathfrak{iso}(1,3)$ arise through well-defined contraction procedures. These parameters $\ell$ and $L$ may be identified respectively with the Planck length $\ell_{\text{P}} = \sqrt{\hbar G / c^3}$ and the de Sitter radius $R_{\text{dS}} = \sqrt{3/\Lambda}$, where $\hbar$ is Planck's constant, $G$ is the gravitational constant, $c$ is the speed of light in vacuum, and $\Lambda$ is the cosmological constant \cite{Ranaivoson2021}.

The paper is organized as follows. Section 2 reviews the notion of Lie algebra contraction from a physical perspective and recalls the standard de Sitter-Poincaré-Galilei chain as a illustrative example. Section 3 analyzes the LCT algebra for one-dimensional space and examines its possible contractions controlled by the two fundamental length parameters $\ell$ and $L$. Section 4 extends the discussion to the relativistic multidimensional case with signature (1,4), where we explicitly derive the contraction structure of the full LCT algebra and identify the conditions under which spacetime symmetry algebras emerge. Section 5 analyzes the emergence of spacetime and momentum space from the geometric structure of the Quantum Phase Space in appropriate contraction limits. Section 6 provides a discussion of the physical implications of these results, their connection to some previous works on quantum phase space symmetry \cite{Ranaivoson2021,Ranaivoson2022,Ranaivoson2025,Randriantsoa2025}, and concluding remarks on the broader significance of this framework for understanding the emergence of spacetime symmetries and for particle and gravitational physics. Concerning the notation, all operators are written in bold. Other notations related to linear algebra are based on the guidelines provided in the reference \cite{Raoelina1985}. Concerning Lie groups and Lie algebras, although we keep the usual mathematical notations, our generators are taken to be Hermitian, which introduces an explicit  factor $
i$ in the commutation relations, in accordance with the physics convention.

\section{Contraction of Lie groups}

\subsection{Definition}

The notion of Lie algebra contraction was introduced by In\"on\"u and Wigner in 1953 as a systematic method for relating two Lie algebras through a limiting procedure acting on their generators \cite{Wigner1953}. 
Contraction plays an important role in mathematical physics, as many physical theories emerge as limiting cases of more general ones, and their corresponding symmetry groups can often be obtained through this mechanism \cite{Celeghini1990,Gilmore2008}.

Typical examples include the Galilei group as a contraction of the Poincar\'e group in the non-relativistic limit $c \to \infty$, and the Poincar\'e group as a contraction of the de Sitter group in the limit where the de Sitter radius $R_{\text{dS}} \to \infty$ \cite{Wigner1953}. Another example is the case of the Carroll group, which is also a contraction of the Poincaré group for $c\rightarrow 0$ \cite{Leblond1965, Leblond2023}.
These examples illustrate how different physical regimes of spacetime symmetry are connected through well-defined limiting processes.

Formally, let $G$ be a Lie group with associated Lie algebra $\mathfrak{g}$ generated by $\{\boldsymbol{X}_i\}$, satisfy
\begin{equation}
        [\boldsymbol{X}_i,\boldsymbol{X}_j] 
=i C_{ij}^{\;\;k} \boldsymbol{X}_k,
\end{equation}
A contraction is defined by introducing a one-parameter family of linear transformations $T_\varepsilon : \mathfrak{g} \to \mathfrak{g}$, invertible for $\varepsilon \neq 0$, such that the transformed generators
\begin{equation}
    \boldsymbol{X}_i'(\varepsilon) = T_\varepsilon(\boldsymbol{X}_i)
\end{equation}
satisfy
\begin{equation}
    [\boldsymbol{X}_i'(\varepsilon),\boldsymbol{X}_j'(\varepsilon)] 
= iC_{ij}^{\;\;k}(\varepsilon) \boldsymbol{X}_k'(\varepsilon),
\end{equation}
where the structure constants depend continuously on $\varepsilon$. 
If the limit
\begin{equation}
    \lim_{\varepsilon \to 0} C_{ij}^{\;\;k}(\varepsilon) 
= \tilde{C}_{ij}^{\;\;k}
\end{equation}
exists, the generators
\begin{equation}
    \tilde{\boldsymbol{X}}_i = \lim_{\varepsilon \to 0} \boldsymbol{X}_i'(\varepsilon)
\end{equation}
define a new Lie algebra $\tilde{\mathfrak{g}}$, called a contraction of $\mathfrak{g}$.

\subsection{Standard examples : De Sitter to Poincar\'e to Galilei}

As a guiding example, consider the chain of contractions
\[
SO(1,4) \longrightarrow ISO(1,3) \longrightarrow \text{Galilei},
\]
which relates the symmetry groups of curved relativistic spacetime, flat relativistic spacetime, and classical non-relativistic spacetime \cite{Bacry1968,Landau1977,Hawking1973}.

The de Sitter group $SO(1,4)$ is generated by ten operators $\boldsymbol{J}_{\alpha\beta}$ satisfying \cite{Weinberg1995}
\begin{equation}
     { [\boldsymbol{J}_{\alpha\beta},\boldsymbol{J}_{\gamma\delta}]}
=
i\hbar(\eta_{\beta\gamma}\boldsymbol{J}_{\alpha\delta}
-\eta_{\alpha\gamma}\boldsymbol{J}_{\beta\delta}
-\eta_{\beta\delta}\boldsymbol{J}_{\alpha\gamma}
+\eta_{\alpha\delta}\boldsymbol{J}_{\beta\gamma}).
\end{equation}
Where $\alpha,\beta=0,1,2,3,4$

Introducing the de Sitter radius $R_{\text{dS}} = \sqrt{3/\Lambda}$ and defining
\begin{equation}
    \boldsymbol{P}_\mu=\frac{1}{R_{\text{dS}}}\boldsymbol{J}_{\mu 4},
\end{equation}
one obtains
\begin{equation}
    [\boldsymbol{J}_{\mu\nu},\boldsymbol{P}_\rho]
=
i\hbar(\eta_{\nu\rho}\boldsymbol{P}_\mu-\eta_{\mu\rho}\boldsymbol{P}_\nu),
\qquad
[\boldsymbol{P}_\mu,\boldsymbol{P}_\rho]
=
\frac{i\hbar}{R_{\text{dS}}^2}\boldsymbol{J}_{\mu\rho}.
\end{equation}
Where $\mu,\nu,\rho=0,1,2,3$

In the flat limit $R_{\text{dS}}\to\infty$, the commutator $[\boldsymbol{P}_\mu,\boldsymbol{P}_\nu]$ vanishes and the algebra reduces to the Poincar\'e algebra 
$\mathfrak{iso}(1,3)=\mathfrak{so}(1,3)\ltimes \mathbb{R}^4$ \cite{Wigner1953,Celeghini1990}.

A further contraction is obtained by introducing the speed of light $c$ and rescaling the boost generators by a factor of $\frac{1}{c}$.
In the limit $c \to \infty$ ($\frac{1}{c} \to 0$), the Poincar\'e algebra reduces to the Galilei algebra which governs Newtonian space and time \cite{Wigner1953,Bacry1968}.

These well-known contractions show how different spacetime symmetry groups describing distinct physical regimes are related through limits controlled by fundamental physical scale parameters such as curvature and the speed of light \cite{Wigner1953,Hall2013}.
In the following sections, a similar mechanism will be applied to the LCT algebra, where the relevant physical scale parameters are the minimum length $\ell$ and the maximum length $L$. These parameters may be naturally associated, respectively, with the Planck length and the de Sitter radius. In this sense, the contraction of the LCT algebra may be viewed as a phase-space extension of the standard de Sitter–Poincar\'e contraction.

\section{LCT group corresponding to one-dimensional space}
\subsection{Definition and associated algebra}

In one dimension, LCTs are defined as linear transformations that mix operators $\boldsymbol{p}_0$ and $\boldsymbol{x}_0$ while preserving the CCR \cite{Ranaivoson2021,Ranaivoson2022,Ranaivoson2025,Randriantsoa2025}. 
They can be written in matrix form as
\begin{equation}
    \begin{pmatrix}
\boldsymbol{p}_0' &
\boldsymbol{x}_0'
\end{pmatrix}
=
\begin{pmatrix}
\boldsymbol{p}_0 & \boldsymbol{x}_0
\end{pmatrix}
\begin{pmatrix}
\mathbb{A} & \mathbb{C} \\
\mathbb{B} & \mathbb{D}
\end{pmatrix}
\end{equation}
with
\begin{equation}
    \begin{cases}
\boldsymbol{p}_0' = \mathbb{A} \boldsymbol{p}_0 + \mathbb{B} \boldsymbol{x}_0, \\
\boldsymbol{x}_0' = \mathbb{C} \boldsymbol{p}_0 + \mathbb{D} \boldsymbol{x}_0,
\end{cases}
\end{equation}
subject to the preservation of the CCR
\begin{equation}
    [\boldsymbol{p}_0',\boldsymbol{x}_0'] =[\boldsymbol{p}_0,\boldsymbol{x}_0]= i\hbar \eta_{00}
\qquad
[\boldsymbol{p}_0',\boldsymbol{p}_0']=[\boldsymbol{p}_0,\boldsymbol{p}_0]=0
\qquad
[\boldsymbol{x}_0',\boldsymbol{x}_0']=[\boldsymbol{x}_0,\boldsymbol{x}_0]=0
\end{equation}
where $\eta_{00}=1$.
The transformation matrix must satisfy
\begin{equation}
    \begin{pmatrix}
\mathbb{A} & \mathbb{C} \\
\mathbb{B} & \mathbb{D}
\end{pmatrix}^{T}
\begin{pmatrix}
0 & \eta_{00} \\
-\eta_{00} & 0
\end{pmatrix}
\begin{pmatrix}
\mathbb{A} & \mathbb{C} \\
\mathbb{B} & \mathbb{D}
\end{pmatrix}
=
\begin{pmatrix}
0 & \eta_{00} \\
-\eta_{00} & 0
\end{pmatrix},
\end{equation}
which implies that the matrix belongs to the symplectic group $Sp(2)$. 
Thus, in one dimension, the LCT group is isomorphic to $Sp(2)$.

It is convenient to introduce the two length scale parameters $\ell$ and $L$ to write the transformations in a dimensionless form:
\begin{equation}
    \begin{cases}
\boldsymbol{p}_0'=\mathbb{A} \boldsymbol{p}_0 + \dfrac{\hbar}{L^2} \mathbb{B} \boldsymbol{x}_0 \\
\boldsymbol{x}_0' =\dfrac{\ell^2}{\hbar} \mathbb{C} \boldsymbol{p}_0 + \mathbb{D} \boldsymbol{x}_0,
\end{cases}
\end{equation}
with $[\boldsymbol{p}_0',\boldsymbol{x}_0']=[\boldsymbol{p}_0,\boldsymbol{x}_0]=i\hbar \eta_{00}=i\hbar$  .

The factors $\dfrac{\hbar}{L^2}$ and $\dfrac{\ell^2}{\hbar}$ compensate for the dimensional mismatch between coordinates and momenta, so that the parameters $\mathbb{A},\mathbb{B},\mathbb{C},\mathbb{D}$ in (13) are dimensionless. 

Since the transformation preserves the CCR, it can also be represented by a unitary operator $U$ acting on the Hilbert space \cite{Weinberg1995,Sakurai2011,Wald1984}:
\begin{equation}
    \boldsymbol{x}_0' = \boldsymbol {U} \boldsymbol{x}_0 \boldsymbol {U}^\dagger,
\qquad
\boldsymbol{p}_0' = \boldsymbol {U} \boldsymbol{p}_0 \boldsymbol {U}^\dagger.
\end{equation}
For infinitesimal transformations \cite{Wigner1959},
\begin{equation}
    \boldsymbol {U} = e^{i\boldsymbol {\theta}} \simeq I + i\boldsymbol {\theta}, \qquad \boldsymbol {U} = e^{-i\boldsymbol {\theta}} \simeq I - i\boldsymbol {\theta}
\end{equation}
where the Hermitian generator $\boldsymbol{\theta}$ can be decomposed on the basis of quadratic operators:
\begin{equation}
    \boldsymbol{\theta} = \boldsymbol {\theta}^1 \boldsymbol{\Omega}^+ + \boldsymbol {\theta}^2 \boldsymbol{\Omega}^- + \boldsymbol {\theta}^3 \boldsymbol{\Omega}^\times.
\end{equation}
The quadratic generators can be defined as 
\begin{equation}
    \begin{cases}
            \boldsymbol{\Omega}^+ =
\frac{\hbar}{2}
\left(
\frac{\ell^2}{\hbar^2} \boldsymbol{p}_0^2
+
\frac{1}{4L^2} \boldsymbol{x}_0^2
\right),\\

\boldsymbol{\Omega}^- =
\frac{\hbar}{2}
\left(
\frac{\ell^2}{\hbar^2} \boldsymbol{p}_0^2
-
\frac{1}{4L^2} \boldsymbol{x}_0^2
\right),\\

\boldsymbol{\Omega}^\times =
\frac{1}{4}(\boldsymbol{p}_0 \boldsymbol{x}_0 + \boldsymbol{x}_0 \boldsymbol{p}_0).
    \end{cases}
\end{equation}
They satisfy the following commutation relations
\begin{equation}
\begin{cases}
    [\boldsymbol{\Omega}^+,\boldsymbol{\Omega}^-]
=
-i\hbar\frac{\ell^2}{L^2}  \boldsymbol{\Omega}^\times\\
[\boldsymbol{\Omega}^-,\boldsymbol{\Omega}^\times]
=
i\hbar \boldsymbol{\Omega}^+\\
[\boldsymbol{\Omega}^\times,\boldsymbol{\Omega}^+]
=
-i\hbar \boldsymbol{\Omega}^-
\end{cases}
\end{equation}

The algebra corresponding to relations (17) and (18) shares some similarities with the dispersion operator algebra defined in \cite {Andriambololona2017}.

In order to elucidate the geometric meaning of the quadratic generators in relation (17), we can evaluate their action on the canonical variables using the fundamental commutation relation $[\boldsymbol{p}_0,\boldsymbol{x}_0]= i\hbar \eta_{00}$ with $\eta_{00}=1$. A direct computation yields
\begin{align}
[\boldsymbol{\Omega}^+,\boldsymbol{x}_0] &= i\ell^2\boldsymbol{p}_0, & [\boldsymbol{x}_0,\boldsymbol{\Omega}^-] &= -i\ell^2\boldsymbol{p}_0, \\
[\boldsymbol{\Omega}^-,\boldsymbol{p}_0] &= -\frac{i\hbar^2}{4L^2}\boldsymbol{x}_0, & [\boldsymbol{\Omega}^+,\boldsymbol{p}_0] &= \frac{i\hbar^2}{4L^2}\boldsymbol{x}_0, \\
[\boldsymbol{\Omega}^\times,\boldsymbol{p}_0] &= -\frac{i\hbar}{2}\boldsymbol{p}_0, & [\boldsymbol{\Omega}^\times,\boldsymbol{x}_0] &= \frac{i\hbar}{2}\boldsymbol{x}_0.
\end{align}
These relations show that $\boldsymbol{\Omega}^+$ and $\boldsymbol{\Omega}^-$ generate the mixing between the momentum and coordinate operators, while $\boldsymbol{\Omega}^\times$ is the generator of dilations.

\subsection{Contraction limits}
\subsubsection{Contraction in the limit \texorpdfstring{$\ell \rightarrow 0$}{l→0}}

To describe the contracted algebra,
we introduce new generators
\begin{equation}
    \boldsymbol{A}_1=\lim_{\ell\to0}\boldsymbol{\Omega}^{+}, \qquad
\boldsymbol{A}_2=\lim_{\ell\to0}\boldsymbol{\Omega}^{-}, \qquad
\boldsymbol{A}_3=\lim_{\ell\to0}\boldsymbol{\Omega}^{\times}
\end{equation}
Taking the limit $\ell \to 0$ in the quadratic generators yields
\begin{equation}
\boldsymbol{A}_1 = \frac{\hbar}{8L^2}\boldsymbol{x}_0^2,\qquad
\boldsymbol{A}_2 = -\frac{\hbar}{8L^2}\boldsymbol{x}_0^2=-\boldsymbol{A}_1,\qquad
\boldsymbol{A}_3 = \boldsymbol{\Omega}^{\times}
\end{equation}
Using the original commutation relations, the contracted algebra becomes
\begin{equation}
[\boldsymbol{A}_1,\boldsymbol{A}_2] = 0,\qquad
[\boldsymbol{A}_2,\boldsymbol{A}_3] = i\hbar \boldsymbol{A}_1,\qquad
[\boldsymbol{A}_3,\boldsymbol{A}_1] = -i\hbar \boldsymbol{A}_2 .
\end{equation}
Since $\boldsymbol{A}_2=-\boldsymbol{A}_1$, only two generators remain independent.
Their commutation relation reads
\begin{equation}
[\boldsymbol{A}_3,\boldsymbol{{A}_1}] = i\hbar \boldsymbol{A}_1 ,
\end{equation}
At the level of the LCT (13), the corresponding explicit expression becomes
\begin{equation}
    \begin{cases}
\boldsymbol{p}_0'= \mathbb{A} \boldsymbol{p}_0 + \dfrac{\hbar}{L^2} \mathbb{B} \boldsymbol{x}_0 \\
\boldsymbol{x}_0' = \mathbb{D} \boldsymbol{x}_0 
\end{cases}
\end{equation}
Thus, for the linear canonical transforms (LCT) governed by the algebra in (22), the new coordinate operator $\boldsymbol{x}_0'$ reduces to a simple scaling of the old one. In contrast, the new momentum operator $\boldsymbol{p}_0'$ must be expressed as a mixture of both the original coordinate and the momentum operators.

\subsubsection{Contraction in the limit \texorpdfstring{$L \rightarrow \infty$}{L→∞}}

The contracted algebra is described by introducing new generators.
\begin{equation}
    \boldsymbol{B}_1=\lim_{L\to\infty}\boldsymbol{\Omega}^{+}, \qquad
\boldsymbol{B}_2=\lim_{L\to\infty}\boldsymbol{\Omega}^{-}, \qquad
\boldsymbol{B}_3=\lim_{L\to\infty}\boldsymbol{\Omega}^{\times}
\end{equation}
In the limit $L \to \infty$, the quadratic generators reduce to
\begin{equation}
\boldsymbol{B}_1 = \frac{\ell^2}{2\hbar}\boldsymbol{p}_0^2,\qquad
\boldsymbol{B}_2 = \frac{\ell^2}{2\hbar}\boldsymbol{p}_0^2=\boldsymbol{B}_1,\qquad
\boldsymbol{B}_3 = \boldsymbol{\Omega}^{\times}
\end{equation}
Applying the original commutation relations, we obtain the contracted algebra.
\begin{equation}
[\boldsymbol{B}_1,\boldsymbol{B}_2] = 0,\qquad
[\boldsymbol{B}_2,\boldsymbol{B}_3] = i\hbar \boldsymbol{B}_1,\qquad
[\boldsymbol{B}_3,\boldsymbol{B}_1] = -i\hbar \boldsymbol{B}_2 .
\end{equation}
Due to the equivalence $\boldsymbol{B}_1 = \boldsymbol{B}_2$, the algebra retains only two independent generators, whose commutation relation is given by
\begin{equation}
[\boldsymbol{B}_3,\boldsymbol{{B}_1}] = -i\hbar \boldsymbol{B}_1 ,
\end{equation}
At the level of the LCT (13), the corresponding explicit expression becomes
\begin{equation}
    \begin{cases}
\boldsymbol{p}_0'=\mathbb{A} \boldsymbol{p}_0  \\
\boldsymbol{x}_0'= \dfrac{\ell^2}{\hbar} \mathbb{C} \boldsymbol{p}_0 + \mathbb{D} \boldsymbol{x}_0,
\end{cases}
\end{equation}
The linear canonical transforms (LCT) corresponding to the algebra defined by relation (27) are therefore transformations for which the new momentum operator $\boldsymbol {p}_0'$ is simply proportional to the old one (dilation or contraction), while the new coordinate operator $\boldsymbol {x}_0'$ is a combination of the old coordinate and momentum operators.

\subsubsection{Contraction in the limit \texorpdfstring{$\ell \rightarrow 0$}{l→0} and \texorpdfstring{$L \rightarrow \infty$}{L→∞}}

In order to present the contracted algebra, we set new generators.
\begin{equation}
    \boldsymbol{C}_1=\lim_{\ell\to0,L\to\infty}\boldsymbol{\Omega}^{+}, \qquad
\boldsymbol{C}_2=\lim_{\ell \to0,L\to\infty}\boldsymbol{\Omega}^{-}, \qquad
\boldsymbol{C}_3=\lim_{\ell\to0,L\to\infty}\boldsymbol{\Omega}^{\times}
\end{equation}
In this simultaneous limit, the quadratic generators become
\begin{equation}
\boldsymbol{C}_1 \rightarrow 0,\qquad
\boldsymbol{C}_2 \rightarrow 0,\qquad
\boldsymbol{C}_3 = \frac14(\boldsymbol{p}_0\boldsymbol{x}_0+\boldsymbol{x}_0\boldsymbol{p}_0).
\end{equation}
In this simultaneous limit, only one generator effectively remains and the resulting algebra reduces to the one–dimensional abelian Lie algebra.

At the level of the LCT (13), the corresponding explicit expression becomes
\begin{equation}
    \begin{cases}
\boldsymbol{p}_0'=\mathbb{A} \boldsymbol{p}_0  \\
\boldsymbol{x}_0' =\mathbb{D}    \boldsymbol{x}_0 =\dfrac{1}{\mathbb{A}}{x}_0
\end{cases}
\end{equation}
 The momentum operator is multiplied by the factor $\mathbb{A}$  , and the coordinate operator is divided by the same factor. This corresponds to a dilation in momentum space and a contraction in coordinate space (if $\lvert \mathbb{A} \rvert >$  1) or vice versa (if $\lvert \mathbb{A} \rvert <$  1).The generator corresponding to this dilation/contraction is $\boldsymbol{C}_3$

\section{Contractions of LCT group for the signature (1,4)}
\subsection{Definition and associated algebra}

We now consider the five-dimensional relativistic quantum phase space with signature $(1,4)$. 
The momenta and coordinate  operators $(\boldsymbol{x}_\alpha,\boldsymbol{p}_\alpha)$, with $\alpha=0,\dots,4$, satisfy \cite{Ranaivoson2021,Ranaivoson2022,Ranaivoson2025,Randriantsoa2025} 
\begin{equation}
    [\boldsymbol{p}_\alpha,\boldsymbol{x}_\beta] = i\hbar \eta_{\alpha\beta},
\qquad
[\boldsymbol{p}_\alpha,\boldsymbol{p}_\beta]=0, 
\qquad
[\boldsymbol{x}_\alpha,\boldsymbol{x}_\beta]=0,
\end{equation}
where
\begin{equation}
    \eta_{\alpha\beta}=\mathrm{diag}(+1,-1,-1,-1,-1).
\end{equation}

The definition of LCT can be written in the form
\begin{equation}
    \begin{cases}
\boldsymbol{p}_\alpha' = \mathbb{A}_\alpha^{\;\beta} \boldsymbol{p}_\beta + \dfrac{\hbar}{L^2} \mathbb{B}_\alpha^{\;\beta} \boldsymbol{x}_\beta, \\
\boldsymbol{x}_\alpha' = \dfrac{\ell^2}{\hbar} \mathbb{C}_\alpha^{\;\beta} \boldsymbol{p}_\beta + \mathbb{D}_\alpha^{\;\beta} \boldsymbol{x}_\beta,
\end{cases}
\end{equation}
with
\begin{equation}
    [\boldsymbol{p}_\alpha',\boldsymbol{x}_\beta']=[\boldsymbol{p}_\alpha,\boldsymbol{x}_\beta] = i\hbar \eta_{\alpha\beta},
\qquad
[\boldsymbol{p}_\alpha',\boldsymbol{p}_\beta']=[\boldsymbol{p}_\alpha,\boldsymbol{p}_\beta]=0, 
\qquad
[\boldsymbol{x}_\alpha',\boldsymbol{x}_\beta']=[\boldsymbol{x}_\alpha,\boldsymbol{x}_\beta]=0,
\end{equation}
The transformation matrix must then satisfy the symplectic condition
\begin{equation}
    \begin{pmatrix}
\mathbb{A} & \dfrac{\ell^2}{\hbar}\mathbb{C} \\
\dfrac{\hbar}{L^2}\mathbb{B} & \mathbb{D}
\end{pmatrix}^{T}
\begin{pmatrix}
0 & \eta \\
-\eta & 0
\end{pmatrix}
\begin{pmatrix}
\mathbb{A} & \dfrac{\ell^2}{\hbar}\mathbb{C} \\
\dfrac{\hbar}{L^2}\mathbb{B} & \mathbb{D}
\end{pmatrix}
=
\begin{pmatrix}
0 & \eta \\
-\eta & 0
\end{pmatrix}.
\end{equation}
For a signature $(1,4)$, this implies that the transformation matrix belongs to the symplectic group $Sp(2,8)$. 
The relativistic LCT symmetry therefore describes a covariant structure acting directly in quantum phase space \cite{Ranaivoson2021,Ranaivoson2022,Ranaivoson2025,Randriantsoa2025}.

$\bold {Remark}:$ As already mentioned in the Introduction, $\ell$ and $L$ may be identified respectively with the Planck length $\ell_{\text{P}} = \sqrt{\hbar G / c^3}$ and the de Sitter radius $R_{\text{dS}} = \sqrt{3 / \Lambda}$, where $\hbar$ is Planck's constant, $G$ is the gravitational constant, and $\Lambda$ is the cosmological constant.

The quadratic hermitian generators corresponding to the unitary representation of the LCT group can be chosen as 
\begin{equation}
    \begin{cases}
        \boldsymbol{\Omega}^+_{\alpha\beta}
=
\frac{\hbar}{2}
\left(
\frac{\ell^2}{\hbar^2}\boldsymbol{p}_\alpha \boldsymbol{p}_\beta
+
\frac{1}{4L^2}\boldsymbol{x}_\alpha \boldsymbol{x}_\beta
\right),\\

\boldsymbol{\Omega}^-_{\alpha\beta}
=
\frac{\hbar}{2}
\left(
\frac{\ell^2}{\hbar^2}\boldsymbol{p}_\alpha \boldsymbol{p}_\beta
-
\frac{1}{4L^2}\boldsymbol{x}_\alpha \boldsymbol{x}_\beta
\right),\\

\boldsymbol{\Omega}^\times_{\alpha\beta}
=
\frac{1}{8}
(\boldsymbol{p}_\alpha \boldsymbol{x}_\beta + \boldsymbol{p}_\beta \boldsymbol{x}_\alpha + \boldsymbol{x}_\alpha \boldsymbol{p}_\beta + \boldsymbol{x}_\beta \boldsymbol{p}_\alpha),\\

\boldsymbol{J}_{\alpha\beta}=\boldsymbol{x}_\alpha \boldsymbol{p}_\beta - \boldsymbol{x}_\beta \boldsymbol{p}_\alpha.
    \end{cases}
\end{equation}
As in the one-dimensional case, $\boldsymbol{\Omega}^+_{\alpha\beta}$ and $\boldsymbol{\Omega}^-_{\alpha\beta}$ represent the mixing of momentum and coordinate operators, while $\boldsymbol{\Omega}^\times_{\alpha\beta}$ and $\boldsymbol{J}_{\alpha\beta}$ generate transformations that mix coordinate operators among themselves or momentum operators among themselves.

They satisfy the following commutation relations
\begin{equation}
    \begin{cases}
        [\boldsymbol{\Omega}^+_{\alpha\beta}, \boldsymbol{\Omega}^-_{\gamma\delta}] = -\frac{i\hbar \ell^2}{4L^2} (\eta_{\alpha\gamma}\boldsymbol{\Omega}^\times_{\beta\delta} + \eta_{\alpha\delta}\boldsymbol{\Omega}^\times_{\beta\gamma} + \eta_{\beta\gamma}\boldsymbol{\Omega}^\times_{\alpha\delta} + \eta_{\beta\delta}\boldsymbol{\Omega}^\times_{\alpha\gamma}) \\
        [\boldsymbol{\Omega}^-_{\alpha\beta}, \boldsymbol{\Omega}^\times_{\gamma\delta}] = \frac{i\hbar}{4} (\eta_{\alpha\gamma}\boldsymbol{\Omega}^+_{\beta\delta} + \eta_{\alpha\delta}\boldsymbol{\Omega}^+_{\beta\gamma} + \eta_{\beta\gamma}\boldsymbol{\Omega}^+_{\alpha\delta} + \eta_{\beta\delta}\boldsymbol{\Omega}^+_{\alpha\gamma}) \\
        [\boldsymbol{\Omega}^\times_{\alpha\beta}, \boldsymbol{\Omega}^+_{\gamma\delta}] = -\frac{i\hbar}{4} (\eta_{\alpha\gamma}\boldsymbol{\Omega}^-_{\beta\delta} + \eta_{\alpha\delta}\boldsymbol{\Omega}^-_{\beta\gamma} + \eta_{\beta\gamma}\boldsymbol{\Omega}^-_{\alpha\delta} + \eta_{\beta\delta}\boldsymbol{\Omega}^-_{\alpha\gamma}) \\
        [\boldsymbol{\Omega}^\times_{\alpha\beta}, \boldsymbol{\Omega}^\times_{\gamma\delta}] = \frac{i\hbar}{2} (\eta_{\alpha\gamma}\boldsymbol{J}_{\beta\delta} + \eta_{\alpha\delta}\boldsymbol{J}_{\beta\gamma} + \eta_{\beta\gamma}\boldsymbol{J}_{\alpha\delta} + \eta_{\beta\delta}\boldsymbol{J}_{\alpha\gamma}) \\
        [\boldsymbol{\Omega}^\times_{\alpha\beta}, J_{\gamma\delta}] = i\hbar (\eta_{\alpha\gamma}\boldsymbol{\Omega}^\times_{\beta\delta} + \eta_{\alpha\delta}\boldsymbol{\Omega}^\times_{\beta\gamma} - \eta_{\beta\gamma}\boldsymbol{\Omega}^\times_{\alpha\delta} - \eta_{\beta\delta}\boldsymbol{\Omega}^\times_{\alpha\gamma}) \\
        [\boldsymbol{J}_{\alpha\beta}, \boldsymbol{\Omega}^+_{\gamma\delta}] = i\hbar (\eta_{\alpha\gamma}\boldsymbol{\Omega}^+_{\beta\delta} + \eta_{\beta\delta}\boldsymbol{\Omega}^+_{\alpha\gamma} - \eta_{\alpha\delta}\boldsymbol{\Omega}^+_{\beta\gamma} - \eta_{\beta\gamma}\boldsymbol{\Omega}^+_{\alpha\delta}) \\
        [\boldsymbol{\Omega}^-_{\alpha\beta}, \boldsymbol{J}_{\gamma\delta}] = i\hbar (\eta_{\alpha\gamma}\boldsymbol{\Omega}^-_{\beta\delta} + \eta_{\alpha\delta}\boldsymbol{\Omega}^-_{\beta\gamma} - \eta_{\beta\gamma}\boldsymbol{\Omega}^-_{\alpha\delta} - \eta_{\beta\delta}\boldsymbol{\Omega}^-_{\alpha\gamma}) \\
    [\boldsymbol{J}_{\alpha\beta},\boldsymbol{J}_{\gamma\delta}]
=i\hbar(\eta_{\beta\gamma}\boldsymbol{J}_{\alpha\delta}
-\eta_{\alpha\gamma}\boldsymbol{J}_{\beta\delta}
-\eta_{\beta\delta}\boldsymbol{J}_{\alpha\gamma}
+\eta_{\alpha\delta}\boldsymbol{J}_{\beta\gamma}).
    \end{cases}
\end{equation}

$\boldsymbol{J}_{\alpha\beta}$ are infinitesimal generators of the de Sitter group $SO(1,4)$, so we may call them de Sitter generators.

\subsection{Contraction limits}

\subsubsection{Contraction in the limit \texorpdfstring{$\ell \rightarrow 0$}{l→0}}

To characterize the contracted algebra, we define the following generators:
\begin{equation}
    \boldsymbol{A}^{+}_{\alpha\beta}
=\lim_{\ell\to0}\boldsymbol{\Omega}^{+}_{\alpha\beta}, \qquad
\boldsymbol{A}^{-}_{\alpha\beta}
=\lim_{\ell\to0}\boldsymbol{\Omega}^{-}_{\alpha\beta}, \qquad
\boldsymbol{A}^{\times}_{\alpha\beta}
=\lim_{\ell\to0}\boldsymbol{\Omega}^{\times}_{\alpha\beta}.
\end{equation}
In the limit $\ell \to 0$, the quadratic generators reduce to
\begin{equation}
    \boldsymbol{A}^{+}_{\alpha\beta}
=
\frac{\hbar}{8L^2}\boldsymbol{x}_\alpha \boldsymbol{x}_\beta,\qquad
\boldsymbol{A}^{-}_{\alpha\beta}
=
-\frac{\hbar}{8L^2}\boldsymbol{x}_\alpha \boldsymbol{x}_\beta=-\boldsymbol{A}^{+}_{\alpha\beta},\qquad
\boldsymbol{A}^{\times}_{\alpha\beta}
=\boldsymbol{\Omega}^{\times}_{\alpha\beta}
\end{equation}
Like the generators $\boldsymbol{A}^{\times}_{\alpha\beta}
=\boldsymbol{\Omega}^{\times}_{\alpha\beta}$, the de Sitter generators $\boldsymbol{J}_{\alpha\beta}=\boldsymbol{x}_\alpha \boldsymbol{p}_\beta-\boldsymbol{x}_\beta \boldsymbol{p}_\alpha$ remain unchanged

Since $\boldsymbol{A}^{+}_{\alpha\beta}=-\boldsymbol{A}^{-}_{\alpha\beta}$, only three types of generator remain independent and the commutators   $[\boldsymbol{A}^{+}_{\alpha\beta},\boldsymbol{A}^{-}_{\gamma\delta}]$ reduce to 0 while all other commutation relations remain unchanged.

At the level of the LCT (37), the corresponding explicit expression becomes
\begin{equation}
    \begin{cases}
\boldsymbol{p}_\alpha' = \mathbb{A}_\alpha^{\;\beta} \boldsymbol{p}_\beta + \dfrac{\hbar}{L^2} \mathbb{B}_\alpha^{\;\beta} \boldsymbol{x}_\beta, \\
\boldsymbol{x}_\alpha' = \mathbb{D}_\alpha^{\;\beta} \boldsymbol{x}_\beta,
\end{cases}
\end{equation}
In the limit $\ell \to 0$, the coordinate operators $\boldsymbol{x}_\alpha$ transform only among themselves through the action of the LCTs generated by the operators $\boldsymbol{A}^{\times}_{\alpha\beta} = \boldsymbol{\Omega}^{\times}_{\alpha\beta}$ and $\boldsymbol J_{\alpha\beta}$, while the new momentum operators $\boldsymbol{p}'_\alpha$ are a linear combination of the original coordinates and momenta.

\subsubsection{Contraction in the limit \texorpdfstring{$L \rightarrow \infty$}{l→0}}

To describe the contracted algebra, we introduce the generators
\begin{equation}
    \boldsymbol{B}^{+}_{\alpha\beta}
=\lim_{L\to\infty}\boldsymbol{\Omega}^{+}_{\alpha\beta}, \qquad
\boldsymbol{B}^{-}_{\alpha\beta}
=\lim_{L\to\infty}\boldsymbol{\Omega}^{-}_{\alpha\beta}, \qquad
\boldsymbol{B}^{\times}_{\alpha\beta}
=\lim_{L\to\infty}\boldsymbol{\Omega}^{\times}_{\alpha\beta}.
\end{equation}
Using the definitions of the quadratic generators we obtain
\begin{equation}
    \boldsymbol{B}^{+}_{\alpha\beta}
=
\frac{\ell^2}{2\hbar}\boldsymbol{p}_\alpha \boldsymbol{p}_\beta,\qquad
\boldsymbol{B}^{-}_{\alpha\beta}
=
\frac{\ell^2}{2\hbar}\boldsymbol{p}_\alpha \boldsymbol{p}_\beta=\boldsymbol{B}^{+}_{\alpha\beta},\qquad
\boldsymbol{B}^{\times}_{\alpha\beta}
= \boldsymbol{\Omega}^{\times}_{\alpha\beta}.
\end{equation}
Both the generators $\boldsymbol{B}^{\times}_{\alpha\beta} = \boldsymbol{\Omega}^{\times}_{\alpha\beta}$ and the de Sitter generators $\boldsymbol{J}_{\alpha\beta} = \boldsymbol{x}_\alpha \boldsymbol{p}_\beta - \boldsymbol{x}_\beta \boldsymbol{p}_\alpha$ remain unchanged.

The condition $\boldsymbol{B}^{+}_{\alpha\beta} = -\boldsymbol{B}^{-}_{\alpha\beta}$ reduces the number of independent generators to three types. As a result, the commutator $[\boldsymbol{B}^{+}_{\alpha\beta}, \boldsymbol{B}^{-}_{\gamma\delta}]$ vanishes, leaving all other commutation relations unchanged.

At the level of the LCT (37), the corresponding explicit expression becomes
\begin{equation}
    \begin{cases}
\boldsymbol{p}_\alpha' = \mathbb{A}_\alpha^{\;\beta} \boldsymbol{p}_\beta \\
\boldsymbol{x}_\alpha' = \dfrac{\ell^2}{\hbar} \mathbb{C}_\alpha^{\;\beta} \boldsymbol{p}_\beta + \mathbb{D}_\alpha^{\;\beta} \boldsymbol{x}_\beta,
\end{cases}
\end{equation}
In the limit $L \to \infty$, the momentum operators $\boldsymbol{p}_\alpha$ transform exclusively among themselves under the action of the LCTs generated by $\boldsymbol{B}^{\times}_{\alpha\beta} = \boldsymbol{\Omega}^{\times}_{\alpha\beta}$ and $\boldsymbol{J}_{\alpha\beta}$. However, the new coordinate operators $\boldsymbol{x}'_\alpha$ become linear combinations of the original coordinates and momenta.

\subsubsection{Contraction in the limit \texorpdfstring{$\ell \rightarrow 0$}{l→0} and \texorpdfstring{$L \rightarrow \infty$}{L→∞}}

To describe the contracted algebra, we introduce the generators
\begin{equation}
\boldsymbol{C}^{+}_{\alpha\beta} = \lim_{\ell \to 0, L \to \infty} \boldsymbol{\Omega}^{+}_{\alpha\beta}, 
\quad 
\boldsymbol{C}^{-}_{\alpha\beta} = \lim_{\ell \to 0, L \to \infty} \boldsymbol{\Omega}^{-}_{\alpha\beta}, 
\quad 
\boldsymbol{C}^{\times}_{\alpha\beta} = \lim_{\ell \to 0, L \to \infty} \boldsymbol{\Omega}^{\times}_{\alpha\beta}.
\end{equation}
Using the definitions of the quadratic generators, we obtain
\begin{equation}
\boldsymbol{C}^{+}_{\alpha\beta} = 0, 
\quad 
\boldsymbol{C}^{-}_{\alpha\beta} = 0, 
\quad 
\boldsymbol{C}^{\times}_{\alpha\beta} = {\Omega}^{\times}_{\alpha\beta}
\end{equation}
Like the generators $\boldsymbol{C}^{\times}_{\alpha\beta}
=\boldsymbol{\Omega}^{\times}_{\alpha\beta}$, the de Sitter generators $\boldsymbol J_{\alpha\beta}=\boldsymbol{x}_\alpha \boldsymbol{p}_\beta-\boldsymbol{x}_\beta \boldsymbol{p}_\alpha$ remain unchanged. This behavior can be understood from the fact that these generators do not explicitly depend on the length scales $\ell$ and $L$, in contrast to $\boldsymbol{C}^{+}_{\alpha\beta}$ and $\boldsymbol{C}^{-}_{\alpha\beta}$, which vanish in the contraction limit.

Although the contraction suppresses part of the algebra, the remaining generators, $\boldsymbol{C}^{\times}_{\alpha\beta}
=\boldsymbol{\Omega}^{\times}_{\alpha\beta}$ and the de Sitter generators $\boldsymbol J_{\alpha\beta}$, still satisfy non-trivial commutation relations inherited from the original algebra. This shows that the contraction does not lead to a completely trivial algebra, but rather to a reduced structure corresponding to the LCTs generated by these remaining generators. In this context, the familiar spacetime symmetries emerge corresponding to the subgroup, of the LCT group, generated by the de Sitter generators. We may distinguish two main cases
\paragraph{Case 1:}

If all de Sitter generators $\boldsymbol{J}_{\alpha\beta}$ are taken to be independent of any contraction parameter, then no new generators arise from the limiting procedure; the corresponding algebra is simply $\mathfrak{so}(1,4)$. Consequently, spacetime translations do not emerge, and the symmetry preserves a de Sitter-type character.

\paragraph{Case 2: Emergence of spacetime translations}

If a non-trivial dependence on the curvature radius, i.e. the de Sitter radius $R_{\text{dS}} = \sqrt{3/\Lambda}$, is instead applied by distinguishing the fifth direction and defining
\begin{equation}
\boldsymbol{P}_\mu = \frac{1}{R_{\text{dS}}} \boldsymbol{J}_{\mu 4}, \qquad \mu = 0,1,2,3,
\end{equation}
then spacetime translations emerge in the contraction limit $R_{\text{dS}} \to \infty$.

Using the $\mathfrak{so}(1,4)$ commutation relations, we compute
\begin{equation}
[\boldsymbol{J}_{\mu\nu}, \boldsymbol{J}_{\rho 4}] 
= i\hbar \left( \eta_{\nu\rho} \boldsymbol{J}_{\mu 4} - \eta_{\mu\rho} \boldsymbol{J}_{\nu 4} \right),
\end{equation}
which implies
\begin{equation}
[\boldsymbol{J}_{\mu\nu}, \boldsymbol{P}_\rho] 
= i\hbar \left( \eta_{\nu\rho} \boldsymbol{P}_\mu - \eta_{\mu\rho} \boldsymbol{P}_\nu \right).
\end{equation}
Moreover,
\begin{equation}
[\boldsymbol{P}_\mu, \boldsymbol{P}_\rho] 
= \frac{1}{R_{\text{dS}}^2} [\boldsymbol{J}_{\mu 4}, J_{\rho 4}] 
= \frac{i\hbar}{R_{\text{dS}}^2} \boldsymbol{J}_{\mu\rho}
\end{equation}
Where $\mu,\nu,\rho=0,1,2,3$

Taking the flat limit $R_{dS} \to \infty$, we obtain
\begin{equation}
\lim_{R_{dS} \to \infty} [\boldsymbol{P}_\mu, \boldsymbol{P}_\rho] = 0.
\end{equation}
The resulting algebra is therefore
\begin{equation}
\mathfrak{iso}(1,3) = \mathfrak{so}(1,3) \ltimes \mathbb{R}^4,
\end{equation}
which is precisely the Poincar\'e algebra.

This shows that spacetime translations arise from the de Sitter sector via a contraction, highlighting the fundamental role of de Sitter generators in the emergence of spacetime symmetry.

As noted previously, the parameter $L$ itself can be identified with the de Sitter radius $R_{\text{dS}} = \sqrt{3/\Lambda}$, though this identification is not mandatory. However, in this identification, spacetime translations emerge directly from the contraction of the LCT group in the limit $L \to \infty$, without any further contraction procedure.

\section{Emergence of spacetime and momentum space from the Quantum Phase Space}

\vspace*{-0.5cm}

For $|\psi\rangle = |\langle \boldsymbol{z}\rangle\rangle$, the expectation values and variance-covariances of the coordinates and momenta operators are \cite{Ranaivoson2021,Ranaivoson2025}:
\begin{equation}
    \begin{cases}
        \langle\boldsymbol{x}_\mu\rangle=\langle\psi|\boldsymbol{x}_\mu|\psi\rangle=\eta_{\mu\nu}\langle\psi|\boldsymbol{x}^\nu|\psi\rangle=\eta_{\mu\nu}\langle\boldsymbol{x}^\nu\rangle\\
        \langle\boldsymbol{p}_\mu\rangle=\langle\psi|\boldsymbol{p}_\mu|\psi\rangle=\eta_{\mu\nu}\langle\psi|\boldsymbol{p}^\nu|\psi\rangle=\eta_{\mu\nu}\langle\boldsymbol{p}^\nu\rangle\\
        \mathcal{X}_{\mu\nu}=\langle\psi|(\boldsymbol{x}_\mu-\langle\boldsymbol{x}_\mu\rangle)(\boldsymbol{x}_\nu-\langle\boldsymbol{x}_\nu\rangle)|\psi\rangle=\eta_{\mu\nu}\mathcal{X}^\rho_\nu\\
        \mathcal{P}_{\mu\nu}=\langle\psi|(\boldsymbol{p}_\mu-\langle\boldsymbol{p}_\mu\rangle)(\boldsymbol{p}_\nu-\langle\boldsymbol{p}_\nu\rangle)|\psi\rangle=\eta_{\mu\nu}\mathcal{P}^\rho_\nu\\
        \mathcal{Q}^\ltimes_{\mu\nu}=\langle\psi|(\boldsymbol{p}_\mu-\langle\boldsymbol{p}_\mu\rangle)(\boldsymbol{x}_\nu-\langle\boldsymbol{x}_\nu\rangle)|\psi\rangle\\
        \mathcal{Q}^\rtimes_{\mu\nu}=\langle\psi|(\boldsymbol{x}_\nu-\langle\boldsymbol{x}_\nu\rangle)(\boldsymbol{p}_\mu-\langle\boldsymbol{p}_\mu\rangle)|\psi\rangle\\
        \mathcal{Q}_{\mu\nu}=\frac{1}{2}(\mathcal{Q}^\ltimes_{\mu\nu}+\mathcal{Q}^\rtimes_{\mu\nu})
    \end{cases}
\end{equation}
For some particular states $|\psi\rangle = |\langle \boldsymbol{z}\rangle\rangle$ corresponding to gaussian-like wavefunctions \cite{Ranaivoson2022,Ranaivoson2025}:
\begin{equation}
    \langle\{\boldsymbol{x}^\mu\}|\{\langle\boldsymbol{z}_\mu\rangle\}=\langle\boldsymbol{x}|\langle\boldsymbol{z}\rangle\rangle= e^{iK}\frac{e^{-\mathcal{B}_{\mu\nu}(\boldsymbol{x}^\mu-\langle\boldsymbol{x}^\mu\rangle)(\boldsymbol{x}^\nu-\langle\boldsymbol{x}^\nu\rangle)-i\langle\boldsymbol{p}_\mu\rangle\boldsymbol{x}^\mu}}{[(2\pi)^D(det\mathcal{X)}]^\frac{1}{4}}
\end{equation}
in which $e^{ik}$ is a unitary complex number that does not depend on $\boldsymbol{x}^\mu$, we have the saturation of the uncertainty relations
\begin{equation}
    \mathcal{P}_{\mu\mu}\mathcal{X}_{\mu\mu}-(\mathcal{Q}_{\mu\mu})^2=\frac{\hbar^2}{4}
\end{equation}
The quantum phase space is defined as the set of possible values 
\(\{\langle \boldsymbol{p}_\mu \rangle, \langle \boldsymbol{x}_\mu \rangle\}\) 
of expectation values for states \(|\boldsymbol{z}\rangle\rangle\), given their 
\(2N \times 2N\) momentum–coordinate variance–covariance matrix
$ 
\begin{pmatrix}
\mathcal{P} & \mathcal{Q} \\
\mathcal{Q}^T & \mathcal{X}
\end{pmatrix}
$ \cite{Ranaivoson2022,Ranaivoson2025}.

Like references \cite{Ranaivoson2021,Ranaivoson2022}, an LCT can be considered a change in the observational frame of reference. Then it is the momenta and coordinates operators that change, but not the state $|\psi\rangle$. Under such transformations, the mean values obey
\begin{equation}
(\langle \boldsymbol{p}' \rangle \ \langle \boldsymbol{x}' \rangle)
=
(\langle \boldsymbol{p} \rangle \ \langle \boldsymbol{x} \rangle)
\begin{pmatrix}
\mathbb{A} & \frac{\ell^2}{\hbar}\mathbb{C} \\
\frac{\hbar}{L^2}\mathbb{B} & \mathbb{D}
\end{pmatrix}.
\end{equation}
If we denotes $\mathcal{P},\mathcal{X}$ and $\mathcal{Q}$ the $N\times N$ matrices corresponding respectively to the mean values and statistical variance-covariances, we can also deduce the following relations
\begin{equation}
    \begin{pmatrix}
        \mathcal{P}' & \mathcal{Q}'\\
    \mathcal{Q}'^T & \mathcal{X}'
    \end{pmatrix}=
    \begin{pmatrix}
        \mathbb{A} & \frac{\ell^2}{\hbar}\mathbb{C}\\
        \frac{\hbar}{L^2}\mathbb{B} & \mathbb{D}
    \end{pmatrix}^T=
    \begin{pmatrix}
        \mathcal{P} & \mathcal{Q}\\
    \mathcal{Q}^T & \mathcal{X}
    \end{pmatrix}
    \begin{pmatrix}
        \mathbb{A} & \frac{\ell^2}{\hbar}\mathbb{C}\\
        \frac{\hbar}{L^2}\mathbb{B} & \mathbb{D}
    \end{pmatrix}
\end{equation}
An important scalar invariant under LCTs is given by
\begin{equation}
\Gamma =
(\langle \boldsymbol{p} \rangle \ \langle \boldsymbol{x} \rangle)
\begin{pmatrix}
\mathcal{P} & \mathcal{Q} \\
\mathcal{Q}^T & \mathcal{X}
\end{pmatrix}^{-1}
(\langle \boldsymbol{p} \rangle \ \langle \boldsymbol{x} \rangle)^T,
\quad
.
\end{equation}
This invariant defines a unified geometric constraint on the quantum phase space, encoding both coordinate and momentum structures in a single covariant relation \cite{Ranaivoson2021,Randriantsoa2026}. If we choose $\Gamma = \frac{L^2}{\ell^2}$, then we can have a transparent physical interpretation in the appropriate contraction limits:
\begin{itemize}
    \item In the limit $\ell \to 0$, the constraint reduces to a relation involving only the coordinate mean values,
\begin{equation}
\langle \boldsymbol{x}_0 \rangle^2 - \sum_{i=1}^{4} \langle \boldsymbol{x}_i \rangle^2 = -L^2,
\end{equation}
which is precisely the equation corresponding to a de Sitter spacetime with radius $R_{dS}=L$.

    \item In the limit $L \to \infty$, one obtains a dual relation in momentum space,
\begin{equation}
\langle \boldsymbol{p}_0 \rangle^2 - \sum_{i=1}^{4} \langle \boldsymbol{p}_i \rangle^2 = -\left(\frac{\hbar}{2\ell}\right)^2,
\end{equation}
corresponding to a de Sitter-like geometry in momentum space.
\end{itemize}

A more direct physical interpretation can be obtained by selecting appropriate reference frames. In particular, choosing a frame in which the spatial components vanish, $\langle \boldsymbol{x}_i \rangle = 0$, or $\langle \boldsymbol{p}_i \rangle = 0$, highlights the role of the fifth components $\langle \boldsymbol{x}_4 \rangle$ and $\langle \boldsymbol{p}_4 \rangle$ as embedding coordinates controlling the curvature scales. In this picture, $L^2$ and $(\hbar/2\ell)^2$ appear as minimal values for $\langle \boldsymbol{x}_4 \rangle^2$ and $\langle \boldsymbol{p}_4 \rangle^2$, respectively.
\begin{itemize}
    \item Finally, in the simultaneous limit $\ell \to 0$ and $L \to \infty$, one recovers the Poincaré regime. Defining
\begin{equation}
\lim_{\ell \to 0,\, L \to \infty} \left(\langle \boldsymbol{x}_4 \rangle^2 - L^2\right) = c^2 \tau^2,
\quad
\lim_{\ell \to 0,\, L \to \infty} \left(\langle \boldsymbol{p}_4 \rangle^2 - \left(\frac{\hbar}{2\ell}\right)^2\right) = c^2 m^2,
\end{equation}
the invariant constraint reduces to the standard relativistic relation
\begin{equation}
\langle \boldsymbol{x}_0 \rangle^2 - \sum_{i=1}^{3} \langle \boldsymbol{x}_i \rangle^2 = c^2 \tau^2,
\quad
\langle \boldsymbol{p}_0\rangle^2 - \sum_{i=1}^{3} \langle \boldsymbol{p}_i \rangle^2 = c^2 m^2,
\end{equation}
\end{itemize}
These final relations are precisely the standard relations of special relativity, where $\tau$ represents the proper time of the particle (i.e., time in its average rest frame) and $m$ denotes its rest mass. As one would expect, the symmetry group associated with both spacetime and momentum space in this context is the Poincaré group.

\section{ Discussion and conclusion}
 The contraction analysis presented in this work clarifies how the familiar symmetry groups of spacetime can arise from the relativistic quantum phase-space symmetry described by the symplectic algebra $\mathfrak{sp}(2,8)$. Linear Canonical Transformations (LCTs), originally developed in signal processing and optics \cite{Wolf1979,Xu2013,Healy2016}, thus reveal themselves as fundamental mathematical tools in relativistic quantum physics \cite{Ranaivoson2021} when suitably generalized to the multidimensional setting dictated by spacetime signature. In this framework, spacetime coordinates and momenta are treated on an equal footing, with spacetime symmetries emerging as effective limits of a more general quantum phase-space structure. This perspective aligns with the idea of an underlying geometry in which conjugate observables are unified \cite{Ranaivoson2022}.
 
A particularly significant aspect of this framework concerns the Coleman-Mandula theorem, which states that the only possible Lie group symmetries of a relativistic interacting theory are direct products of the Poincaré group and an internal symmetry group \cite{Coleman1967,Oskar1997}. As discussed in references \cite{Ranaivoson2021} and \cite{Ranaivoson2025}, the LCT symmetry framework may circumvent this restriction because it acts on quantum phase space rather than spacetime alone.

The contraction procedure is controlled by two fundamental length scale parameters $\ell$ and $L$, which may be identified respectively with the Planck length and the de Sitter radius. As demonstrated in Sections 3 and 4, different combinations of these parameters yield contracted algebras leading to the de Sitter algebra $\mathfrak{so}(1,4)$ and, in the flat limit, the Poincaré algebra $\mathfrak{iso}(1,3)$. This mechanism, analogous to the Inönü-Wigner contraction \cite{Wigner1953,Celeghini1990,Gilmore2008}, appears naturally within the broader context of quantum phase-space symmetry, with the de Sitter generators $\boldsymbol{J}_{\alpha\beta}$  persisting throughout all contraction limits associated with $\ell$ and $L$. However, when the de Sitter radius $R_{\text{dS}}$ is introduced and taken to infinity, the de Sitter generators contract to the spacetime translations and Lorentz generators of the Poincaré group. This additional contraction may merge with those associated with $L$ if it is identified with the de Sitter radius.

Recent investigations reveal direct connections to particle physics. References \cite{Ranaivoson2021,Ranaivoson2022,Ranaivoson2025,Randriantsoa2025} demonstrate that the spin representation of the LCT group for signature $ (1,4) $ allows the establishment of a new classification of quarks and leptons, suggesting the existence of sterile neutrinos. These developments can be connected to fundamental questions in gravitation, particle physics, and cosmology \cite{Ranaivoson2025,Randriantsoa2025,Aldrovandi2007,Sakurai2011,Wald1984,Kibble1961,Weinberg1972,Govaerts1990}, where the parameters $\ell$  and $L$ may naturally  invoke the Planck scale and the cosmological constant.

The geometric analysis in Section~5 reveals that the LCT-invariant 
$\Gamma = L^2/\ell^2$ provides a common framework to describe both 
coordinate and momentum geometries. Taking the contraction $\ell \to 0$ 
leads to a de Sitter spacetime, while $L \to \infty$ gives a de Sitter-like momentum space. The combined limit $\ell \to 0$ and  $ L \to \infty$ recovers the 
familiar Poincaré-invariant relations corresponding to the definition of proper time $\tau$ and rest mass $m$ as shown in relation (65).

In conclusion, the symplectic structure underlying relativistic LCTs offers a unified framework in which spacetime symmetries emerge as effective structures from a deeper quantum phase-space geometry through well-defined contraction mechanisms. By operating at a level more fundamental than spacetime itself, this framework may naturally circumvent no-go theorems such as Coleman-Mandula while offering promising connections to both gravitational physics and the internal symmetries observed in particle physics \cite{Ranaivoson2025,Randriantsoa2025}. This perspective, reminiscent of Born's reciprocity principle \cite{Born1949, Castro2025}, suggests that the symmetries of spacetime may ultimately find their origin in a more primitive quantum phase-space geometry

\newpage

\bibliographystyle{unsrt}
\bibliography{references}

\end{document}